\documentclass[11pt]{article}
\pdfoutput=1
\usepackage{sint,amsfonts}
\usepackage[T1]{fontenc}
\usepackage[utf8]{inputenc}
\usepackage{amssymb}
\usepackage{amsmath}
\usepackage{simplewick}
\usepackage{booktabs}
\usepackage{siunitx}
\usepackage{physics}
\usepackage{wrapfig}
\usepackage[caption=false]{subfig}
\usepackage{hyperref}

\newcommand{\tableref}[1]{Table~\ref{#1}}
\newcommand{\figref}[1]{Figure~\ref{#1}}

\newcommand{\be}{\begin{equation}}
\newcommand{\ee}{\end{equation}}
\newcommand{\bea}{\begin{eqnarray}}
\newcommand{\eea}{\end{eqnarray}}

\newcommand{\lalg}[1]{\mathfrak{#1}}
\newcommand{\ad}[1]{\text{ad}_{#1}}

\newcommand{\Vd}{V^\dagger}

\newcommand{\fm}{\femto\meter}

\newcommand{\dmu}{\hat{\mu}}
\newcommand{\dnu}{\hat{\nu}}

\begin{document}

\begin{titlepage}

\begin{center}
\begin{flushright}
\end{flushright}
\vspace{1.0cm}

{\Large\bf Non-Gaussianities in the topological charge distribution\\[0.15cm]
of the SU(3) Yang--Mills theory\\[0.5ex]} 

\end{center}
\vskip 0.5 cm
\begin{center}
{\large  Marco C\`e$^{\scriptscriptstyle a}$, Cristian Consonni$^{\scriptscriptstyle b}$, 
Georg P. Engel$^{\scriptscriptstyle c,}$\footnote{Present address: AEE INTEC, Feldgasse 19, 
8020 Gleisdorf, Austria.} and Leonardo Giusti$^{\scriptscriptstyle c}$}
\vskip 0.75cm
$^{\scriptstyle a}$ Scuola Normale Superiore, Piazza dei Cavalieri 7, 56126 Pisa, Italy\\
and INFN, Sezione di Pisa, Largo B. Pontecorvo 3, 56127 Pisa, Italy\\
\vskip 0.25cm
$^{\scriptstyle b}$ Dipartimento di Ingegneria e Scienza dell'Informazione, Universit\`a di Trento\\
\vskip 0.25cm
$^{\scriptstyle c}$ Dipartimento di Fisica, Universit\`a di Milano-Bicocca, and INFN,\\
Sezione di Milano-Bicocca, Piazza della Scienza 3, 20126 Milano, Italy\\ 
                
\vskip 2.0cm
{\bf Abstract}
\vskip 0.35ex
\end{center}

\noindent
We study the topological charge distribution of the 
SU(3) Yang--Mills theory with high precision in order to be 
able to detect deviations from Gaussianity. The computation 
is carried out on the lattice with high statistics Monte Carlo
simulations by implementing a naive discretization of the topological 
charge evolved with the Yang--Mills gradient flow. This definition
is far less demanding than the one suggested from Neuberger's 
fermions and, as shown in this paper, in the continuum limit its 
cumulants coincide with those of the universal definition appearing 
in the chiral Ward identities. 
Thanks to the range of lattice volumes and spacings considered, we 
can extrapolate 
the results for the second and fourth cumulant of the topological 
charge distribution to the continuum limit with confidence by keeping 
finite volume effects negligible with respect to the statistical errors. 
Our best results for the topological susceptibility is 
$t_0^2\, \chi=\num{6.67(7)e-4}$, where $t_0$ is a standard reference 
scale, while for the ratio of the forth cumulant over the second 
we obtain $R=0.233(45)$. The latter 
is compatible with the expectations from the large $N_c$ expansion, 
while it rules out the $\theta$-behavior of the vacuum energy predicted 
by the dilute instanton model. Its large distance from 1 implies 
that, in the ensemble of gauge configurations that dominate the 
path integral, the fluctuations of the topological charge are of 
quantum {\it non-perturbative} nature. 
\vskip 4.5cm

\vfill

\eject

\end{titlepage}

\section{Introduction}
The discovery of a fermion operator~\cite{Neuberger:1997bg} 
that satisfies the Ginsparg--Wilson (GW) relation~\cite{Ginsparg:1981bj} 
triggered a breakthrough in our understanding of the topological 
effects in Quantum Chromodynamics (QCD) and in the Yang--Mills 
theory~\cite{Hasenfratz:1998ri,Luscher:1998pqa,Giusti:2001xh,Giusti:2004qd,Luscher:2004fu}.
This progress made it possible to give a precise and unambiguous 
implementation of the Witten--Veneziano 
formula~\cite{Witten:1979vv,Veneziano:1979ec,Giusti:2001xh,Seiler:2001je}. 

In lattice QCD a naive definition
of the topological charge density needs to be combined with an unambiguous 
renormalization condition. The cumulants of the charge,
for instance the susceptibility, require also additional subtractions of 
short-distance singularities to make them integrable distributions. If the 
topological charge density is defined as suggested by GW 
fermions~\cite{Neuberger:1997bg,Hasenfratz:1998ri,Luscher:1998pqa}, however,   
its bare lattice expression and those of the corresponding cumulants have finite and unambiguous 
continuum limits as they stand, which in turn satisfy the anomalous 
chiral Ward identities~\cite{Giusti:2001xh,Giusti:2004qd,Luscher:2004fu}. 
By combining a series of those identities, the cumulants can be written as integrated 
correlation functions of scalar and pseudoscalalar density chains or combination of 
them~\cite{Giusti:2004qd,Luscher:2004fu,Giusti:2008vb}. In this form a particular
regularization is not required anymore to prove that no renormalization factor or 
subtractions of short-distance singularities are required. These expressions provide 
a universal definition of the susceptibility and of the higher cumulants which 
satisfy the anomalous chiral Ward Identities~\cite{Luscher:2004fu}.

Recently a new definition of the topological charge was 
found~\cite{Luscher:2010iy}, whose cumulants have a finite and 
unambiguous continuum limit~\cite{Luscher:2010iy,Luscher:2011bx}. It is a naive 
discretization of the charge evolved with the Yang--Mills gradient flow. 
It is particularly appealing because its
numerical evaluation is significantly cheaper than the one for the definition 
suggested by Neuberger's fermions. Here we show that in the Yang--Mills theory 
the cumulants defined this way coincide, in the continuum limit, with those of the 
universal definition appearing in the anomalous chiral Ward Identities of QCD. 
By implementing the gradient-flow definition, we compute the topological susceptibility 
in the continuum limit with a precision 5 times better than the reference computation 
with the Neuberger's definition~\cite{DelDebbio:2004ns}. We then determine 
the ratio of the forth cumulant over the second one in the continuum limit 
by keeping for the first time all systematics, especially finite volume effects, negligible with 
respect to the statistical errors. As a byproduct we also perform an interesting 
universality test at the permille level by comparing the values of the topological 
susceptibility at different flow times.

\section{Preliminaries in the continuum\label{sec:cont}}
Starting from the ordinary fundamental gauge field 
\be
B_\mu \Big|_{t=0} = A_\mu\; , 
\ee
where $A_\mu = A^a_\mu T^a$ (see Appendix~\ref{ap:conventions} for the generator conventions), 
the Yang--Mills gradient flow evolves the gauge field as a function
of the flow time $t\geq 0$ by solving the differential
equation~\cite{Luscher:2010iy} 
\bea
\partial_t B_\mu & = & D_\nu G_{\nu\mu} + \alpha_0 D_\mu \partial_\nu B_\nu\; , \\[0.25cm]
G_{\mu\nu} & = & \partial_\mu B_\nu - \partial_\nu B_\mu - i[B_\mu,B_\nu]\; , \qquad
D_\mu = \partial_\mu - i\, [B_\mu,\cdot]\; ,
\eea
with $\alpha_0$ being the parameter which determines the gauge. Here we focus on 
the gradient-flow evolution of the topological charge density defined 
as\footnote{If not explicitly indicated, the superscript $t$ on the 
quantities evolved with the gradient flow is always $>0$.} 
\be\label{qtcont}
q^t = \frac{1}{32 \pi^2} \epsilon_{\mu\nu\rho\sigma} \tr\Big[G_{\mu\nu} G_{\rho\sigma}\Big] \; , 
\ee
and of the corresponding topological charge
\be
Q^t = \int d^4 x\, q^t(x)\; ,  
\ee
where $\epsilon_{\mu\nu\rho\sigma}$ is the four-index totally antisymmetric tensor and 
the trace is over the color index. Under a generic variation $\delta B_\mu$ of a given 
gauge field configuration, the topological charge density changes as 
\be\label{eq:poly}
\delta q^t = \partial_\rho \tilde w^t_\rho\; , \qquad 
 \tilde w^t_\rho = \frac{1}{8 \pi^2} \epsilon_{\rho\mu\nu\sigma} 
\tr\left[G_{\mu\nu}\, \delta B_\sigma \right]\; , 
\ee
see for instance Ref.~\cite{Polyakov:1987ez}. If we now specify 
\be
\delta B_\mu = \partial_t B_\mu\, \delta t \; ,  
\ee
it is straightforward to show that  
\be\label{eq:polyWF}
\partial_t q^t = \partial_\rho w^t_\rho\; , \qquad 
w^t_\rho = \frac{1}{8 \pi^2} \epsilon_{\rho\mu\nu\sigma} 
\tr\left[G_{\mu\nu} D_\alpha G_{\alpha\sigma} \right]\; , 
\ee
where $w^t_\rho$ is a local dimension-$5$ gauge-invariant pseudovector field.
This in turn implies that for a given gauge field configuration 
\be\label{eq:polyWF2}
\partial_t Q^t = 0\; ,  
\ee 
an equation which reflects the topological nature of $Q^t$.\\[-0.325cm]

When $q^t(x)$ is inserted in a correlation function, 
Eq.~(\ref{eq:polyWF}) implies
\be\label{eq:lhsexp}
\langle q^t(x)\, O(y) \rangle = \langle q^{t=0}(x)\, O(y) \rangle + 
\partial_\rho \int_0^t dt'\, \langle w^{t'}_\rho(x)\, O(y)\rangle
\qquad (x \neq y)\; ,
\ee
where $O(y)$ is any finite (multi)local operator inserted at a physical 
distance from $x$. The l.h.s. of Eq.~(\ref{eq:lhsexp}) is finite 
thanks to the fact that a gauge-invariant local composite 
field constructed with the gauge field evolved at positive flow time 
is finite~\cite{Luscher:2010iy,Luscher:2011bx}. 
Since there are no local composite fields of dimension $d<5$ with the symmetry properties of 
$w^t_{\rho}(x)$, the integrand on the r.h.s 
of Eq.~(\ref{eq:lhsexp}) diverges at most logarithmically when $t'\rightarrow 0$. This 
implies that the quantity
\be\label{eq:qt0def}
\langle q^{t=0}(x)\, O(y) \rangle \equiv \lim_{t\rightarrow 0}\, \langle q^t(x)\, O(y) \rangle \qquad (x \neq y)\; ,   
\ee
is finite, i.e. the limit on the r.h.s exists for any finite operator $O(y)$. The 
Eq.~(\ref{eq:qt0def}) can be taken as the definition of $q^{t=0}(x)$, i.e. the renormalized 
topological charge density operator at $t=0$. The latter satisfies the proper singlet chiral Ward 
identities when fermions are included, see next section for an explicit derivation.
It is worth noting that Eq.~(\ref{eq:lhsexp}) implies that the small-$t$ 
expansion of $q^t(x)$ is of the form 
\be\label{eq:smallt}
\langle q^t(x)\, O(y) \rangle = 
\langle q^{t=0}(x)\, O(y) \rangle + {\cal O}(t) \qquad (x \neq y)\; .   
\ee

In the following we will be interested in the 
cumulants of the topological charge 
\be\label{eq:polyWF3}
{C}^t_{n} = \int d^4 x_1 \dots d^4 x_{2n-1} \langle q^t(x_1)\dots q^t(x_{2n})\rangle_\mathrm{c}\; , 
\ee
which, thanks to Eq.~(\ref{eq:polyWF2}), are expected to be independent of the flow-time
for $t\geq 0$ with the limit $t\rightarrow 0$ which requires some care due to the possible 
appearance of short-distance singularities. It is the aim of the next section
to address this question by using the lattice, a regularization where the
theory can be non-perturbatively defined. We will supplement the
theory with extra degenerate valence quarks of mass $m$, and we will consider the
(integrated) correlator of a topological charge density with a chain made of scalar and
pseudoscalar densities~\cite{Luscher:2004fu} defined as  
\be\label{eq:chain_cont}
\hspace{-0.325cm}  
\langle q^{t=0}(0) P_{51}(z_1) S_{12}(z_2) S_{23}(z_3) S_{34}(z_4) S_{45}(z_5) \rangle\; , 
\ee
where $S_{ij}$ and $P_{ij}$ are the scalar and the pseudoscalar renormalized densities with
flavor indices $i$ 
and $j$. Power counting and the operator product expansion predict that there are no
non-integrable short-distance singularities when the coordinates of two or
more densities in (\ref{eq:chain_cont}) tend to coincide among themselves
or with $0$. When only one of the densities is close to $q^{t=0}(0) $,
the operator product expansion predicts the leading singularity to be  
\be\label{eq:leadingsds}
q^{t=0}(x)\, S_{ij}(0) \xrightarrow{x\to\, 0}
c(x)\, P_{ij}(0) + \dots
\ee
where $c(x)$ is a function which diverges as 
$|x|^{-4}$ when $|x|\rightarrow 0$, and the dots indicate 
sub-leading contributions. An analogous expression 
is valid for the pseudoscalar density. Being the leading short-distance singularity 
in the product of fields $q^{t=0}(x)\, S_{ij}(0)$, 
its Wilson coefficient $c(x)$  can be computed in perturbation theory. By using 
Eq.~(\ref{eq:lhsexp}), to all orders in perturbation
theory\footnote{Since the function $|x|^{-4} \ln(x^2)^{-p}$ is integrable for
$p>1$, the singularity needs to be determined only up to some finite
order.}
we can write
\be\label{eq:bellissima}
\langle q^{t=0}(x) S_{ij}(0)\, O(y) \rangle = 
\langle q^{t}(x) S_{ij}(0)\, O(y) \rangle - 
\partial_\rho \int_0^t dt' 
\langle w^{t'}_{\rho}(x) S_{ij}(0) \, O(y) \rangle \; ,
\ee
where again $O(y)$ is any finite (multi)local operator inserted at a physical 
distance from $0$ and $x$. When $t>0$, the first member on the r.h.s 
of Eq.~(\ref{eq:bellissima}) has no singularities when $|x|\rightarrow 0$.
If present, the singularity has to come from the second term, and therefore
$c(x)$ must be of the form 
\be\label{eq:bellissima2}
c(x) = \partial_\rho u_\rho(x)
\ee
which does not contribute to the integral (over all coordinates) of the 
correlation function (\ref{eq:chain_cont}).

\section{Cumulants of the topological charge on the lattice\label{sec:Qlat}}
On the lattice the Yang--Mills gradient-flow equation can be written as a 
first-order differential equation~\cite{Luscher:2010iy}
\begin{equation}
\label{eq:gradient_flow_compact1}
\partial_t  V_\mu(x) = - g_0^2 \{ \partial_{x,\mu} S(V)\}V_\mu(x), \qquad V_\mu(x)|_{t=0} 
= U_\mu(x)\; , 
\end{equation}
where the Wilson action $S$ and the link differential operators $\partial_{x,\mu}$ are 
defined in Appendix~\ref{ap:conventions} together with other conventions. 
The gauge field evolved at positive flow-time $V_\mu(x)$
is smooth on the scale of the cut-off. When inserted at a physical distance, 
the gauge-invariant local composite fields constructed with the evolved gauge field
are finite as they stand. Remarkably their universality class is determined only by their 
asymptotic behavior in the classical continuum limit~\cite{Luscher:2010iy,Luscher:2011bx}.
At $t>0$ any decent definition of the topological charge density is 
therefore finite. The same line of 
argumentation applies to the cumulants of the topological charge. At $t>0$ short-distance
singularities cannot arise because of the  exponential damping of the high-frequency 
components of the fields enforced by the flow evolution. 

It remains to be shown, however, that the cumulants of the topological charge distribution defined 
at $t>0$ satisfy the proper singlet chiral Ward identities when fermions are included 
in the theory. To show this it is sufficient to work 
with a particular discretization of the topological charge, and then appeal to the 
above mentioned universality argument for the other definitions. The GW discretizations 
have a privileged r\^ole since at $t=0$ the lattice bare cumulants are finite, 
and they satisfy the singlet chiral Ward identities when 
fermions are included in the theory. 

\subsection{Ginsparg--Wilson definition of the charge density}
The definition of the topological charge density suggested 
by GW fermions is~\cite{Neuberger:1997fp,Luscher:1998pqa,Hasenfratz:1998ri}
\begin{equation}\label{eq:qx}
a^4 q^t_\textsc{n}(x) = -\frac{\bar a}{2}\, \tr\Big[\gamma_5 D(x,x)\Big] , 
\end{equation}
where we indicate it with a subscript $\textsc{n}$ since, for concreteness,
we take $D(x,y)$ to be the Neuberger--Dirac operator 
given in Appendix~\ref{ap:conventions} in which each link variable 
$U_\mu(x)$ is replaced by the corresponding evolved one $V_\mu(x)$ when $t>0$. 
Since there are no other operators of dimension $d\leq 4$ which are 
pseudoscalar and gauge-invariant, it holds that 
\be\label{eq:zq1}
\lim_{a\rightarrow 0} Z_q\, \langle q^t_\textsc{n}(0)\, q^{t=0}_\textsc{n}(x) \rangle = 
{\rm finite}\; ,  
\ee
where $Z_q$ is a renormalization constant which is at most logarithmically divergent,
while $q^t_\textsc{n}(0)$ is finite as it stands. This in turn implies that 
\be\label{eq:cd1}
\lim_{a\rightarrow 0} Z_q\, a^4 \sum_x\, \langle q^t_\textsc{n}(0)\, q^{t=0}_\textsc{n}(x) 
\rangle = {\rm finite}\; ,
\ee
since there are no short-distance singularities that contribute to the integrated 
correlation function because $q^t_\textsc{n}(0)$ is evolved at positive flow-time. 
By supplementing the theory  with extra 
degenerate valence quarks of mass $m$, and by replacing in Eq.~(\ref{eq:cd1}) 
the topological charge at $t=0$ with its density-chain 
expression~\cite{Luscher:2004fu} we obtain
\be\label{eq:chain1}
\!\!\!
a^4\! \sum_x \langle q^{t}_\textsc{n}(0)\, q^{t=0}_\textsc{n}(x)\rangle\!  = -m^5 a^{20}\!\!\!\!
\sum_{z_1,\dots,z_5}\!\!\!\!  
\langle q^{t}_\textsc{n}(0)\, P_{51}(z_1)\, S_{12}(z_2)\, S_{23}(z_3)\, S_{34}(z_4)\, S_{45}(z_5) 
\rangle\; , 
\ee
where $S_{ij}$ and $P_{ij}$ are the scalar and the pseudoscalar densities with flavor indices $i$ 
and $j$. 
Written as in Eq.~(\ref{eq:chain1}), power counting and the operator product expansion predict 
that there are no non-integrable short-distance singularities when the coordinates 
of two or more densities tend to coincide. The r.h.s of Eq.~(\ref{eq:chain1}) is finite as 
it stands, and it converges to the continuum limit with a rate proportional to $a^2$.
This in turn implies that the limits
on the l.h.s of Eqs.~(\ref{eq:zq1}) and (\ref{eq:cd1}) are reached with the same rate
if $Z_q$ is set to any fixed ($g_0$-independent) value. Since in the classical continuum
limit Neuberger's definition 
in Eq.~(\ref{eq:qx}) has the same asymptotic behavior of the definition in 
Eq.~(\ref{qtcont}) \cite{Kikukawa:1998pd,Fujikawa:1998if}, we may set
$Z_q=1$ in which case
\be\label{eq:bellaq}
\lim_{a\rightarrow 0}\, \langle q^{t=0}_\textsc{n}(x)\, O_\textsc{l}(y) \rangle  =
\langle q^{t=0}(x)\, O(y) \rangle \qquad (x \neq y)\; ,
\ee
where $O_\textsc{l}(y)$ is a discretization of the generic finite continuum operator $O(y)$. 
Once inserted in correlation functions at a physical distance from 
other (renormalized) fields, $q^{t=0}_\textsc{n}(x)$ does not require any renormalization 
in the Yang--Mills theory. It is finite as it stands, and it satisfies the singlet Ward 
identities when fermions are included in the theory. It is interesting to note that 
Eqs.~(\ref{eq:smallt}) and (\ref{eq:bellaq}) implies 
\be\label{eq:shtfte}
\langle q^{t}(x)\, O(y) \rangle =   \langle q^{t=0}_\textsc{n}(x)\, O_\textsc{l}(y) \rangle +
{\cal O}(a^2) + {\cal O}(t)\; , 
\ee
where in general discretization effects depend on $t$. 
We could have arrived to Eq.~(\ref{eq:bellaq}) by following a procedure analogous 
to the one in the continuum, see Eqs.~(\ref{eq:polyWF})--(\ref{eq:smallt}).
To all orders in perturbation theory, or in general when Neuberger's operator is differentiable 
with respect to the gauge field~\cite{Hernandez:1998et}, the change of the topological charge 
density with respect to the flow-time can be 
written, analogously to Eq.~(\ref{eq:polyWF}), 
as~\cite{Luscher:1998kn,Luscher:2000zd} (see also \cite{Luscher:2000hn})  
\be\label{eq:polyWFN}
\partial_t q^t_\textsc{n}(x) = \partial^{*}_\rho w^t_{\textsc{n}, \rho}(x)\; , 
\ee
where $w^t_{\textsc{n}, \rho}(x)$ is a discretization of the dimension-$5$ gauge-invariant 
pseudovector operator $w^t_{\rho}(x)$, and $\partial^{*}_\rho$ is the backward 
finite-difference operator. 

\subsection{Ginsparg--Wilson definition of the charge cumulants}
The Neuberger's definition of the topological charge is given by 
\be
Q^t_\textsc{n} \equiv a^4 \sum_x q^t_\textsc{n}(x)\; , 
\ee  
and its cumulants are defined as 
\be\label{eq:CnQCd}
{C}^t_{\textsc{n},n} = a^{8n-4}\!\!\!\!\!\!\sum_{x_1,\dots,x_{2n-1}}
\langle q^t_\textsc{n}(x_1)\dots q^t_\textsc{n}(x_{2n-1})\, q^t_\textsc{n}(0)\rangle_\mathrm{c} \; .
\ee
For $t=0$ the cumulants have an unambiguous universal continuum limit as they stand and,
when fermions are included, they satisfy the proper singlet chiral 
Ward identities~\cite{Giusti:2001xh,Giusti:2004qd,Luscher:2004fu}.
They are the proper quantities to be inserted in the Witten--Veneziano 
relations for the mass and scattering amplitudes of the $\eta'$ meson 
in QCD~\cite{Witten:1979vv,Veneziano:1979ec,Giusti:2001xh,Seiler:2001je}.
It is far from being obvious that ${C}^{t=0}_{\textsc{n},n}$ coincide with those defined at 
positive flow-time, since the two definitions may differ by additional finite 
contributions from short-distance singularities.

For the clarity of the presentation we start by focusing on the lowest cumulant, 
the topological susceptibility ${C}^{t}_{\textsc{n},1}$. At $t=0$, by replacing one 
of the two $q^{t=0}_\textsc{n}$ 
with its density-chain expression~\cite{Luscher:2004fu}, we obtain
\be\label{eq:chain}
\hspace{-0.325cm}
a^4\! \sum_x \langle q^{t=0}_\textsc{n}(0) q^{t=0}_\textsc{n}(x)\rangle\!  =\! -m^5 a^{20}\!\!\!\!\!
\sum_{z_1,\dots,z_5}\!\!\!\!  
\langle q^{t=0}_\textsc{n}(0) P_{51}(z_1) S_{12}(z_2) S_{23}(z_3) S_{34}(z_4) S_{45}(z_5) \rangle\; . 
\ee
When the susceptibility is written in this form, the discussion toward the end of
section \ref{sec:cont} and in particular Eq.~(\ref{eq:bellissima2}) guarantee that
there are no contributions from short-distance singularities. This result, together with
the fact that $Z_q=1$, implies that  
\be\label{eq:chit0}
\lim_{t\rightarrow 0}\, \lim_{a\rightarrow 0} a^4 \sum_x \langle q^{t}_\textsc{n}(x)\, q^{t=0}_\textsc{n}(0)\rangle = 
\lim_{a\rightarrow 0} a^4 \sum_x \langle q^{t=0}_\textsc{n}(x)\, q^{t=0}_\textsc{n}(0)\rangle \; .  
\ee
By replacing on the l.h.s $q^{t=0}_\textsc{n}(0)$ with the evolved one, no further 
short-distance singularities are introduced and we arrive to the final result
\be\label{eq:finalres}
\lim_{t\rightarrow 0}\, \lim_{a\rightarrow 0} a^4 \sum_x 
\langle q^{t}_\textsc{n}(x)\, q^{t}_\textsc{n}(0)\rangle = 
\lim_{a\rightarrow 0} a^4 \sum_x \langle q^{t=0}_\textsc{n}(x)\, q^{t=0}_\textsc{n}(0)\rangle \; .
\ee

By replacing $2n-1$ of the charges in the $n^{th}$ cumulant with their density-chain 
definitions, the very same line of argumentation can be applied. The 
Eq.~(\ref{eq:finalres}), together with the independence up to harmless discretization effects 
of ${C}^t_{\textsc{n},n}$ from the flow-time for $t>0$~\cite{Luscher:2010iy},
implies that the continuum limit of 
${C}^t_{\textsc{n},n}$ coincides with the one of ${C}^{t=0}_{\textsc{n},n}$. The cumulants
of the topological charge distribution defined at $t>0$ thus 
satisfy the proper singlet chiral Ward identities when fermions are 
included~\cite{Giusti:2001xh,Giusti:2004qd,Luscher:2004fu}.
They are the proper quantities to be inserted in the Witten--Veneziano relations 
for the mass and scattering amplitudes of the $\eta'$ meson in 
QCD~\cite{Witten:1979vv,Veneziano:1979ec,Giusti:2001xh,Seiler:2001je}.

\subsection{Universality at positive flow-time \label{sec:naive}}
For $t>0$ different lattice definitions of the topological charge density 
belong to the same universality class if they share the same asymptotic 
behavior in the classical 
continuum limit~\cite{Luscher:2010iy,Luscher:2011bx}. In the rest of this 
paper we are interested in the naive 
definition of the topological charge density defined 
as\footnote{We use the same notation for the naive definition of the field strength tensor, 
of the topological charge and of its density 
on the lattice and in the continuum, since any ambiguity is resolved from the context.} 
\begin{equation}
\label{eq:topological_charge}
q^t(x) = \frac{1}{64\pi^2}\, 
\epsilon_{\mu\nu\rho\sigma}\, G_{\mu\nu}^a(x) G_{\rho\sigma}^a(x)\; ,
\end{equation}
where the field strength tensor $G^a_{\mu\nu}(x)$ is defined 
as~\cite{Caracciolo:1991cp}
\begin{equation}
\label{eq:field_strength_tensor_clover}
G_{\mu\nu}^a(x) = - \frac{i}{4 a^2}
\tr[\left( Q_{\mu\nu}(x) - Q_{\nu\mu}(x) \right)\, T^a ]\; , 
\end{equation}
with 
\bea
    Q_{\mu\nu}(x) & = &\, V_\mu(x) V_\nu(x + a\dmu) \Vd_\mu(x + a\dnu) \Vd_\nu(x) + \nonumber\\[0.125cm]
               & &\, V_\nu(x) \Vd_\mu(x - a\dmu + a\dnu) \Vd_\nu(x - a\dmu) V_\mu(x - a\dmu) +\nonumber\\[0.125cm]
 & &\, \Vd_\mu(x - a\dmu) \Vd_\nu(x - a\dmu - a\dnu) V_\mu(x - a\dmu - a\dnu) V_\nu(x - a\dnu) + \\[0.125cm]
 & &\, \Vd_\nu(x - a\dnu) V_\mu(x - a\dnu) V_\nu(x + a\dmu - a\dnu) \Vd_\mu(x)\; .\nonumber     
\eea
In the Yang--Mills theory $q^{t=0}(x)$ requires a multiplicative renormalization
constant\footnote{This renormalization constant can be fixed by enforcing 
the analogous of Eqs.~(\ref{eq:zq1}) and (\ref{eq:bellaq}).} when inserted in correlation 
functions at a physical distance from other 
operators \cite{Alles:1996nm}. The cumulants of the corresponding topological 
charge $Q^{t=0} \equiv a^4 \sum_x q^{t=0}(x)$, defined analogously to 
Eq.~(\ref{eq:CnQCd}), have additional ultraviolet power-divergent singularities, 
and they do not have a well defined continuum limit. 

The density $q^{t}(x)$ in 
Eq.~(\ref{eq:topological_charge}) shares with $q^{t}_\textsc{n}(x)$  
the same asymptotic behavior in the classical 
continuum limit~\cite{Kikukawa:1998pd,Fujikawa:1998if}.
Since for $t>0$ short-distance singularities cannot arise, 
$C^t_{\textsc{n},n}$ and $C^t_{n}$ tend to the same continuum limit. 
The results in the previous section then imply that the continuum limit of the 
naive definition of $C^t_n$, {\it at positive flow-time}, coincides with the 
universal definition which satisfies the chiral Ward identities when fermions 
are added~\cite{Giusti:2001xh,Giusti:2004qd,Luscher:2004fu}. It is interesting to note, however, that 
at fixed lattice spacing there can be quite some differences. For instance, the topological 
susceptibility defined at $t>0$ with the naive definition is not guaranteed to go to zero in the 
chiral limit at finite lattice spacing in presence of fermions~\cite{Bruno:2014ova}. 

\section{Numerical setup}
For the numerical computation  we discretize the SU(3) Yang--Mills theory with the standard Wilson plaquette 
action on a finite four-dimensional lattice with spacing $a$, with the same $L/a$ 
size in all four space-time directions, and with periodic boundary conditions imposed 
on the gauge fields, see Appendix~\ref{ap:conventions} for details. 
The basic Monte Carlo update of each link variable implements 
the Cabibbo--Marinari scheme~\cite{Cabibbo:1982zn}, by sweeping the full lattice
with one heatbath update followed by $L/(2a)$ sweeps of over-relaxation updates. 
\begin{table}[tb]
  \centering
  \begin{tabular}{cS[table-format=1.2]S[table-format=1.2]S[table-format=2]S[table-format=1.1]S[table-format=1.3]S[table-format=7]S[table-format=3]S[table-format=1.3]S[table-format=1.4]S[table-format=1.4]}
    \toprule
    Lattice   & {$\beta$} & {$t_0/a^2$} & {$L/a$} & {$L$\,[\si{\fm}]} & {$a$\,[\si{\fm}]} & {$N_\text{conf}$} & {\texttt{nit}} & {\texttt{eerr}} & {\texttt{q2err}} & {\texttt{q4err}} \\
    \midrule
    $A_1$ &    5.96   &  2.79  &    10   &  1.0  &  0.102  &      36000       &  30       & 0.19 & 0.0005 & 0.0024 \\
    $B_1$ &           &        &    12   &  1.2  &         &     144000       &           & 0.45 &        & 0.005  \\
    $C_1$ &           &        &    13   &  1.3  &         &     280000       &           & 0.42 &        & 0.0068 \\
    $D_1$ &           &        &    14   &  1.4  &         &     505000       &           & 0.74 &        & 0.01   \\
    $E_1$ &           &        &    15   &  1.5  &         &     880000       &           & 0.89 &        & 0.012  \\
    $F_1$ &           &        &    16   &  1.6  &         &    1440000       &           & 1.04 &        & 0.015  \\
    \midrule
    $B_2$ &    6.05   &  3.78  &    14   &  1.2  &  0.087  &     144000       &  60       & 0.31 & 0.0005 & 0.005  \\
    $D_2$ &           &        &    17   &  1.5  &         &     144000       &           & 0.045&        & 0.01   \\
    \midrule
    $B_3$ &    6.13   &  4.87  &    16   &  1.2  &  0.077  &     144000       &  90       & 0.25 & 0.0005 & 0.005  \\
    $D_3$ &           &        &    19   &  1.5  &         &     144000       &           & 0.058&        & 0.01   \\
    \midrule
    $B_4$ &    6.21   &  6.20  &    18   &  1.2  &  0.068  &     144000       & 250       & 0.20 & 0.0005 & 0.005  \\
    $D_4$ &           &        &    21   &  1.4  &         &     144000       &           & 0.042&        & 0.01   \\
    \bottomrule
  \end{tabular}
  \caption{Overview of the ensembles and statistics used in this study. For each lattice we give the 
          label, $\beta=6/g_0^2$, the reference scale $t_0/a^2$, the spatial extent of 
          the lattice, the lattice spacing, the number $N_\text{conf}$ of independent configurations generated, the number of sweeps
          {\texttt{nit}} required to space them, and the tolerances \texttt{eerr}, \texttt{q2err} and  \texttt{q4err} on the primary 
          observables considered (see main text).}
  \label{tab:simulation_details}
\end{table}

\subsection{Ensembles generated}
We have simulated three series of lattices in order 
to estimate and remove the systematic effects due to the finiteness 
of the lattice spacing and volume, see \tableref{tab:simulation_details}
for details. In the first series $\{A_1, B_1,\dots, F_1\}$ the inverse 
coupling $\beta=6/g_0^2$ is kept fixed so that the 
lattice spacing is approximatively $0.1$~fm, while the physical volume increases 
from $(\SI{1.0}{\fm})^4$ to $(\SI{1.6}{\fm})^4$. The number $N_\text{conf}$  
of independent gauge configurations generated scales with $L^8$ to ensure that the relative 
statistical error on $R$, the ratio of the fourth over the second cumulant of the 
topological charge distribution see Eq.~(\ref{eq:obs}), is always at the $10\%$ 
level~\cite{Giusti:2007tu}. In the second series $\{B_1,\dots, B_4\}$ the physical 
volume is kept approximatively fixed, while the 
spacing is decreased down to \SI{0.068}{\fm}. The volume is always 
$(\SI{1.2}{\fm})^4$ to guarantee that finite-size effects on $R$ 
are within the statistical errors, while the computational cost remains affordable. 
In the third series $\{D_1,\dots, D_4\}$ is again the physical 
volume which is kept approximatively fixed, always at least $(\SI{1.4}{\fm})^4$, to 
guarantees that finite-size effects in the reference scale $t_0$ and in the topological susceptibility 
$\chi$, see  Eq.~(\ref{eq:obs}), are within their (smaller) statistical errors. In both cases 
the measurements at the four lattice spacings are used to estimate discretization effects in the 
observables, and to extrapolate them away in the continuum limit.

\subsection{Computation of the observables}
The primary observables that we have computed on each configuration at $t\geq 0$ are the 
energy density
\begin{equation}
    E^t =  \frac{a^4}{4V} \sum_x F_{\mu\nu}^{a,t}(x) F_{\mu\nu}^{a,t}(x)\; ,
\end{equation}
and the topological charge $Q$ defined as in section~\ref{sec:naive}. The quantum 
averages we are interested in are
\be\label{eq:obs}
\langle E^t \rangle\; , \qquad 
\chi^t = \frac{\langle [Q^{t}]^2 \rangle }{V}\; ,\qquad
R^t = \frac{\langle [Q^{t}]^4 \rangle_c}{\langle [Q^{t}]^2 \rangle}\; . 
\ee
To numerically integrate the Yang--Mills gradient flow we have 
implemented a fourth order \emph{Runge--Kutta--Munthe-Kaas (RKMK) method}~\cite{Munthe-Kaas:1995,Munthe-Kaas:1998,Munthe-Kaas:1999}.
It is a \emph{structure-preserving} Runge--Kutta (RK) integrator, designed to exactly preserve the Lie group structure 
of the gradient flow equation, see Appendix~\ref{ap:runge_kutta} for details.
On each lattice the field has been evolved approximatively up to $t=1.2\, t_0$, where $t_0$
is the reference flow-time value defined below.
The observables in Eq.~(\ref{eq:obs}) have been computed with a flow-time resolution of $0.08 a^2$ or smaller. The numerical integration of the flow equation introduces a
\begin{wrapfigure}{r}{7.5cm}
  \centering
  \includegraphics{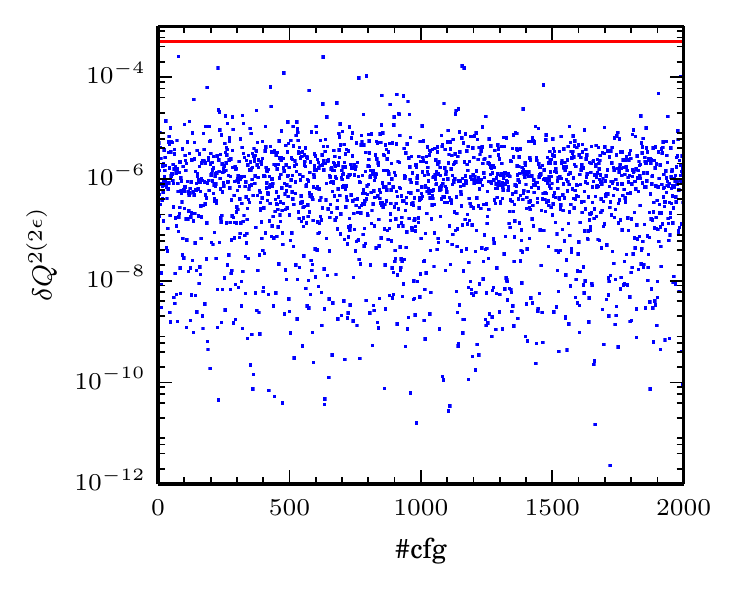}
  \caption{History plot of the systematic error $\delta Q^{2, (2\epsilon)}$ at 
   $t\simeq t_0$ for the first $2000$ configurations of $D_4$. The red line indicates the bound 
   $\mathtt{q2err}=0.0005$ of the systematic error enforced on all configurations.}
  \label{fig:systematicQ2}
\end{wrapfigure}
 systematic error in the gauge field values at 
positive flow-time $t$, and thus in each observable. In our case, at asymptotically small values of the RK 
step size, it is proportional to $\epsilon^4$. There are, however, large fluctuations in
the pre-factor among the various gauge configurations, see Figure~\ref{fig:systematicQ2}. A reliable estimate of 
this systematics is achieved 
by monitoring the error configuration by configuration, and occasionally adapt the step $\epsilon$. To do so 
we integrate the flow equation two times with steps $\epsilon$ and $2 \epsilon$, where in our case 
$\epsilon=0.08 a^2$. Denoting with $E_j^{(\epsilon)}$ and 
$Q_j^{(\epsilon)}$ the basic observables $E$ and $Q$ respectively computed on the 
$j^\text{th}$ field configuration 
evolved with step size $\epsilon$, at small enough $\epsilon$ the error is given by 
\vspace{0.375cm}

\begin{equation}
\label{eq:Qsyst_2eL}
\delta E_j^{(2\epsilon)} = \left| E_j^{(\epsilon)} - E_j^{(2\epsilon)} \right|,\qquad  
\delta Q_j^{(2\epsilon)} = \left| Q_j^{(\epsilon)} - Q_j^{(2\epsilon)} \right|,
\end{equation}
with both observables obviously measured at the same flow-time. By applying linear propagation,
the error on the average over all configurations is bounded by
\bea
& &\!\!\!\!\!\!\!\!\!\!\!\! 
\delta \bar E_j^{(2\epsilon)} \leq \max_j \Big(\delta E_j^{(2\epsilon)}\Big)\,,\;
  \delta \bar{Q}^{2, (2\epsilon)} \leq \max_j \left( |2Q_j| \delta Q_j^{(2\epsilon)} \right)\, ,\;
  \delta \bar{Q}^{4, (2\epsilon)} \leq \max_j \left( |4Q^3_j| \delta Q_j^{(2\epsilon)} \right)\, ,\nonumber\\[0.25cm]
& & \delta \bar R^{(2\epsilon)} \leq \frac{1}{\bar{Q}^2} \max \left( \max_j
\left(|4Q^3_j| \delta Q_j^{(2\epsilon)} \right), \frac{\bar{Q}^4+3(\bar{Q}^2)^2}{\bar{Q}^2} 
\max_j\left(|2Q_j| \delta Q_j^{(2\epsilon)}\right) \right)\; . 
\eea
At run-time, for each configuration and each flow-time the systematic errors of the observables
$E$, $Q^2$ and $Q^4$ are compared with the given tolerances \texttt{eerr}, \texttt{q2err} and 
\texttt{q4err} respectively. If one of the tests fails, the flow evolution is re-computed for that 
configuration with a new step size 
$\epsilon'=(1/2)\epsilon$ and new observables data, along with old $\epsilon=2\epsilon'$ data, are used to 
estimate the systematic errors and compare them with the tolerances. If the test fails again, the field 
is evolved with $\epsilon''=(1/2)\epsilon'$, and so on. This ensures 
that 
\begin{equation}
\label{eq:finalerr}
\delta \bar E_j^{(2\epsilon)} \leq \mathtt{eerr}\; , \quad
  \delta \bar{Q}^{2, (2\epsilon)} \leq \mathtt{q2err}\; , \quad 
\delta \bar{Q}^{4, (2\epsilon)}\leq \mathtt{q4err}\; .
\end{equation}
The parameters \texttt{eerr}, \texttt{q2err} and \texttt{q4err} are chosen as a function of the 
target statistical error on the corresponding observables. If we set the upper limit for the 
systematic error to be roughly 10 times smaller than the statistical one, 
this condition is readily translated into a limit for \texttt{eerr}, \texttt{q2err} and \texttt{q4err}, 
see \tableref{tab:simulation_details} for the values chosen for each lattice. 
The Eqs.~\eqref{eq:finalerr} put bounds

\begin{wrapfigure}{r}{7.5cm}
  \centering
\vspace{-0.5cm}

  \includegraphics{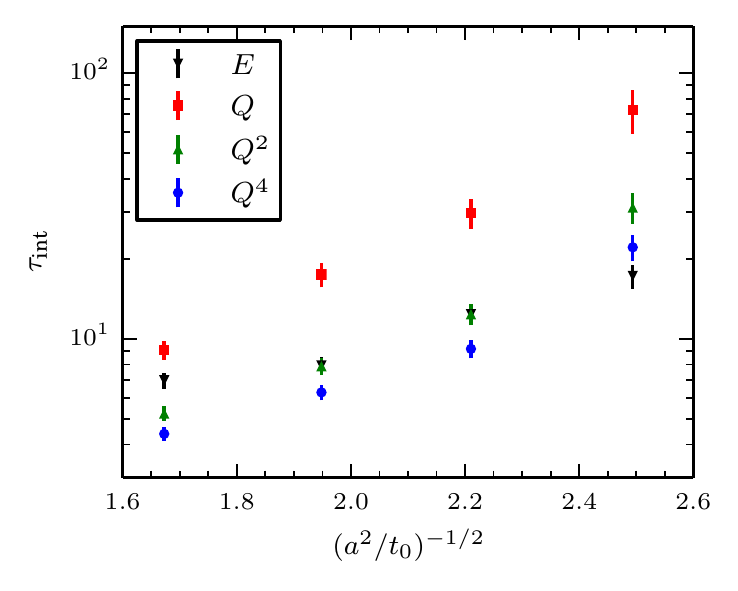}
  \caption{The integrated autocorrelation times $\tau_\text{int}$ of the primary observables as a function 
of $(a^2/t_0)^{-1/2}$.}
  \label{fig:autocorr}
\vspace{-1.5cm}

\end{wrapfigure}

\noindent  on the systematic errors 
for the coarser evolution, but the data evolved with the finer step size $\epsilon$ are 
actually those used in the final analysis. 
This choice is rather conservative in our case,
being the actual error more than one order of magnitude smaller. 
For the quantity $E^t$ the actual error 
turns out to be more than 
two orders of magnitude smaller with respect to the bound in Eq.~(\ref{eq:finalerr}), 
see Figure~\ref{fig:RK_comparison} in Appendix~\ref{ap:runge_kutta}.
We have therefore chosen
larger values for $\mathtt{eerr}$ with respect to one given by the bound in Eq.~(\ref{eq:finalerr}).    
\vspace{0.5cm}

\subsection{Autocorrelation times}
To measure the autocorrelation time of the various observables, we perform a dedicated run 
for each lattice $\{B_1,\dots, B_4\}$ where the gauge field configurations are separated 
by a single iteration of the update algorithm. Each series is replicated $36$ times to increase 
statistical accuracy. The integrated autocorrelation times $\tau_\text{int}$ of the observables 
$E$, $Q$, $Q^2$, and $Q^4$, estimated as in Ref.~\cite{Wolff:2003sm}, are reported in \tableref{tab:autocorrelation}.
\begin{table}[tb]  
  \centering
  \begin{tabular}{ccS[table-format=1.2]S[table-format=2.1(2)]S[table-format=2.1(2)]S[table-format=2.1(2)]S[table-format=2.2(2)]}
    \toprule
    Lattice   & {$N_\text{conf}$} & {$t/a^2$} & {$\tau_\text{int}^E$} & {$\tau_\text{int}^Q$} & {$\tau_\text{int}^{Q^2}$} & {$\tau_\text{int}^{Q^4}$} \\ 
    \midrule
    $B_{1a}$ &  $36 \times 1000$ & 3.36 &  7.0(5)  &  9.1(7)  &  5.2(3)  &  4.39(25) \\ 
    $B_{2a}$ &  $36 \times 1000$ & 4.64 &  7.9(6)  & 17.4(18) &  7.9(6)  &  6.3(4) \\
    $B_{3a}$ &  $36 \times 1000$ & 6.08 & 12.4(11) & 30(4)    & 12.4(11) &  9.2(7) \\
    $B_{4a}$ &  $36 \times 1000$ & 7.68 & 17.2(18) & 73(13)   & 31(4)    &  22.1(25)  \\
    \bottomrule
  \end{tabular}
\caption{Integrated autocorrelation times of the various observables in units of a single sweep 
of the update algorithm. They have been measured on dedicated runs made of $36$ series of $1000$ sweeps
each.\label{tab:autocorrelation}}
\end{table}
In the range of $\beta$ values considered, $Q$ has the largest autocorrelation time which  
increases rapidly toward the continuum limit~\cite{DelDebbio:2002xa}. To ensure that the measurements 
in the main runs are statistically independent, we have spaced them by {\texttt{nit}} sweeps of 
the lattice, see \tableref{tab:simulation_details}.

\section{Physics results}
\begin{figure}[t]
  \centering
  \includegraphics{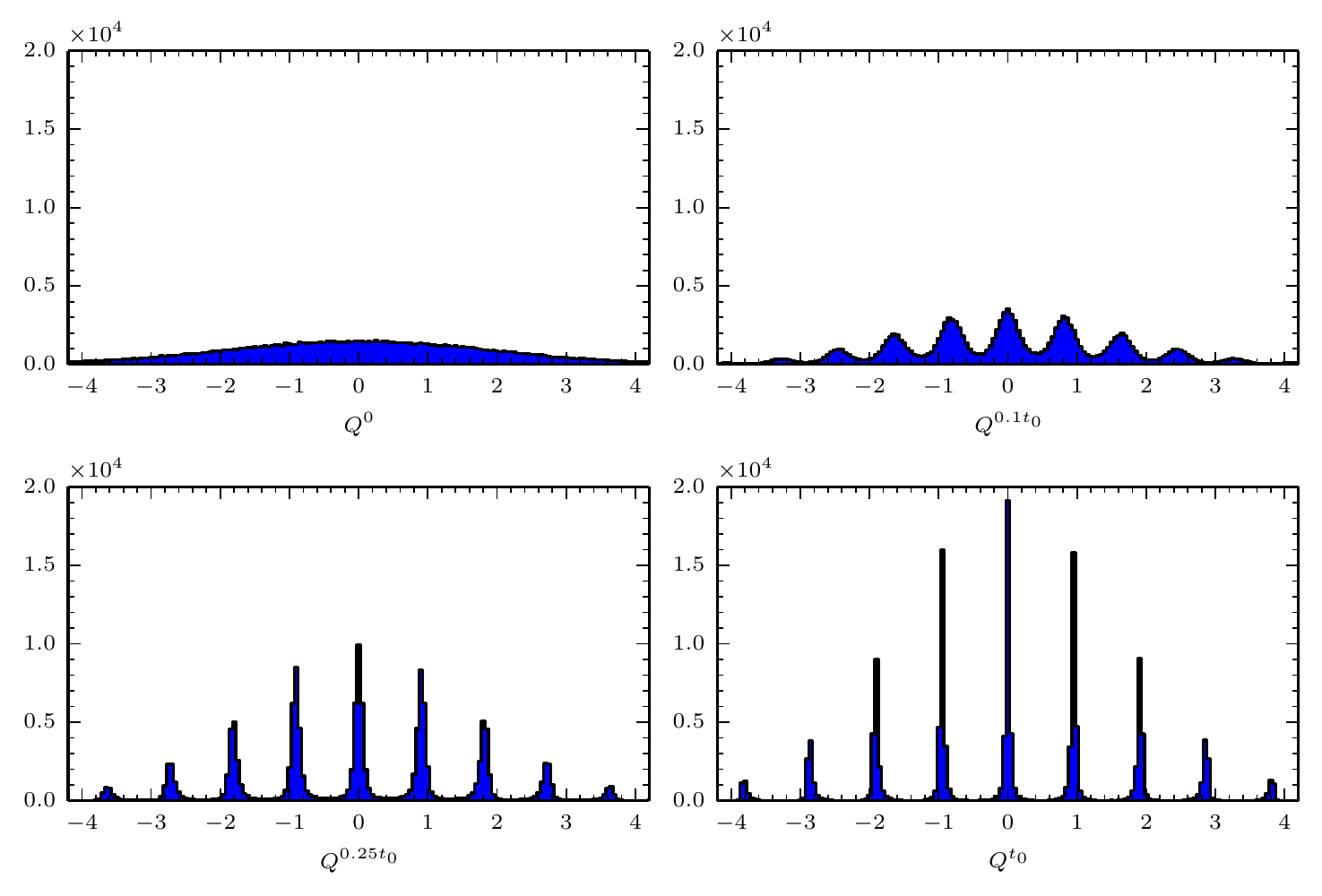}
  \caption{Histograms of the topological charge distribution measured on the lattice $D_4$ 
at different flow-times.}
  \label{fig:distribution_plot}
\end{figure}

A first analysis of the data reveals the effectiveness of the gradient flow in 
splitting the field space of the lattice theory into different topological sectors. In 
\figref{fig:distribution_plot} we plot the histograms of the topological charge $Q$ measured 
at different flow-times on the lattice $D_4$. In the plot on the top-left corner, the 
topological charge distribution at $t=0$ is a smooth function over non-integer values. By 
increasing the flow-time, the configurations with charge close to integers become more 
and more probable. The spikes in the bottom-right plot turn out to be slightly shifted 
towards zero with respect to the integer values due to discretization effects. On the 
other lattices similar histograms are obtained.

\subsection{Scale setting}
\label{sec:scale_setting_t0}
The reference flow-time $t_0$ is defined through the implicit equation~\cite{Luscher:2010iy}
\begin{equation}
\label{eq:t0_def}
  \left. t^2 \expval{E^t} \right|_{t=t_0} = 0.3\, . 
\end{equation}
In the region of interest $t^2\expval{E^t}$ grows approximatively as a linear function of $t$. 
Since we have computed $\expval{E^t}$ at flow-times spaced by finite steps, 
we have solved equation~\eqref{eq:t0_def} by interpolating linearly the two 
data points closest to $t_0$. The results are reported in 
\tableref{tab:t0_results}, with the systematic error due to the interpolation
being negligible. By comparing the values of $t_0/a^2$ obtained on the lattices $\{A_1,\dots, F_1\}$, 
finite-size effects are not visible at the level of $0.1$ permille in the statistical 
precision for $L\geq 1.4$~fm. We thus fix the lattice spacing at all 
values of $\beta$ from $t_0/a^2$ determined on the lattices $\{D_1,\dots, D_4\}$. 
\begin{table}[t]
\begin{center}
  \begin{tabular}{cS[table-format=1.5(2)]S[table-format=1.4(2)]}
    \toprule
    Lattice & {$t_0/a^2$} & {$t_0/r_0^2$} \\
    \midrule
    $A_1$ & 2.995(4)    &   0.1195(9)    \\[0.0325cm]
    $B_1$ & 2.7984(9)   &   0.1117(9)    \\[0.0325cm]
    $C_1$ & 2.7908(5)   &   0.1114(9)    \\[0.0325cm]
    $D_1$ & 2.7889(3)   &   0.1113(9)    \\[0.0325cm]
    $E_1$ & 2.78892(23) &   0.1113(9)    \\[0.0325cm]
    $F_1$ & 2.78867(16) &   0.1113(9)    \\[0.0325cm]
    \bottomrule
  \end{tabular}\qquad
  \begin{tabular}{cS[table-format=1.5(2)]S[table-format=1.4(2)]}
    \toprule
    Lattice & {$t_0/a^2$} & {$t_0/r_0^2$} \\
    \midrule
    $B_2$ & 3.7960(12)  &   0.1114(9)    \\
    $B_3$ & 4.8855(15)  &   0.1113(10)   \\
    $B_4$ & 6.2191(20)  &   0.1115(11)   \\
    \midrule
    $D_2$ & 3.7825(8)   &   0.1110(9)    \\
    $D_3$ & 4.8722(11)  &   0.1110(10)   \\
    $D_4$ & 6.1957(14)  &   0.1111(11)   \\
    \bottomrule
  \end{tabular}
\caption{Results for the reference flow-time $t_0/a^2$ and the 
ratio $t_0/r_0^2$. The error on the latter is dominated by the
\SIrange[range-phrase=--,range-units=single]{0.3}{0.6}{\percent} relative error
on $r_0/a$ quoted in~\cite{Guagnelli:1998ud}.\label{tab:t0_results}}
\end{center}
\end{table}

\begin{wrapfigure}{R}{7.5cm}
  \centering
  \includegraphics{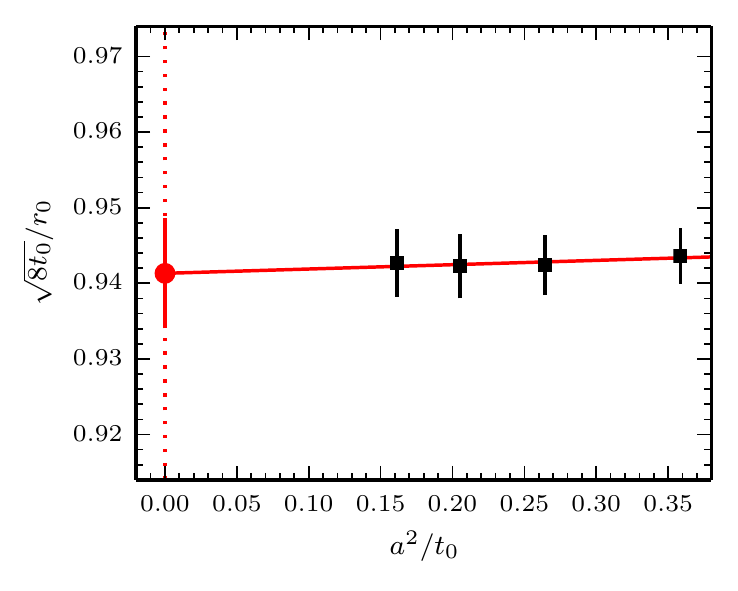}
  \caption{Continuum limit extrapolation of $\sqrt{8t_0}/r_0$ computed on the lattices $\{D_1,\dots, D_4\}$. The errors are dominated by the \SIrange[range-phrase=--,range-units=single]{0.3}{0.6}{\percent} relative error on $r_0/a$ quoted in~\cite{Guagnelli:1998ud}.}
  \label{fig:t0r0}
\end{wrapfigure}

In \tableref{tab:t0_results} the values of $t_0/r_0^2$, where $r_0$ is the Sommer scale
computed in~\cite{Guagnelli:1998ud}, are also reported. As shown in Figure~\ref{fig:t0r0},
discretization effects in this ratio are indeed negligible with respect to the statistical 
errors dominated by the \SIrange[range-phrase=--,range-units=single]{0.3}{0.6}{\percent} 
error on $r_0/a$. By extrapolating the results linearly in 
$a^2/t_0$, we obtain in the continuum limit
\be\label{eq:t0r0}
\sqrt{8t_0}/r_0 = \num{0.941(7)}\, ,  
\ee
which corresponds to $t_0/r_0^2 = \num{0.1108(17)}$. 
To express $t_0$ in physical units,
we supplement the theory with quenched quarks. The value of 
$F_K r_0=0.293(7)$ from Ref.~\cite{Garden:1999fg} together with 
$F_K=109.6$~MeV leads to\footnote{Note that we use an updated determination
for the physical value of $F_K$ with respect to Ref.~\cite{Garden:1999fg}. The
change on $F_K r_0$ induced by the new tuning of the strange quark mass is negligible
with respect to the statistical error quoted.} 
\begin{equation}
\label{eq:t0fm_continuum_limit}
  t_0 = \left(\SI{0.176(4)}{\femto\meter}\right)^2\, , 
\end{equation}
the error being dominated by the one on $F_K r_0$.

\subsection{Topological susceptibility}
The full set of results for the topological charge moments and cumulants 
are given in \tableref{tab:results}. They are computed\footnote{Unless explicitly indicated, 
the gradient flow-time at which the topological quantities are computed throughout this and 
the next sections is $t=t_0$.} at the reference flow-time $t_0$ by linearly interpolating 
the numerical data as described in the previous section. 
\begin{table}[t]
  \centering
  \begin{tabular}{cS[table-format=1.3(2)]S[table-format=2.2(2)]S[table-format=1.3(2)]S[table-format=1.3(2)]}
    \toprule
      Lattice    & {$\langle Q^2\rangle$} & {$\langle Q^4\rangle$} & {$\langle Q^4\rangle_c$} &  {$R$}  \\
    \midrule
    $A_1$ &    0.701(6)   &    1.75(4)    &        0.273(20)       & 0.39(3) \\  
    $B_1$ &    1.617(6)   &    8.15(7)    &        0.30(4)       & 0.187(24) \\ 
    $C_1$ &    2.244(6)   &   15.50(10)   &        0.40(5)       & 0.177(23) \\
    $D_1$ &    3.028(6)   &   28.14(14)   &        0.63(7)       & 0.209(23) \\
    $E_1$ &    3.982(6)   &   48.38(18)   &        0.81(9)       & 0.202(23) \\
    $F_1$ &    5.167(6)   &   80.90(22)   &        0.81(11)      & 0.157(22) \\
    \midrule
    $B_2$ &    1.699(7)   &    9.07(9)    &        0.41(5)       & 0.24(3) \\
    $D_2$ &    3.686(14)  &   41.6(4)     &        0.83(19)      & 0.22(5) \\
    \midrule
    $B_3$ &    1.750(7)   &    9.58(9)    &        0.39(5)       & 0.22(3) \\
    $D_3$ &    3.523(13)  &    37.8(3)    &        0.56(17)      & 0.16(5) \\
    \midrule
    $B_4$ &    1.741(7)   &    9.44(9)    &        0.35(5)       & 0.20(3) \\
    $D_4$ &    3.266(12)  &    32.7(3)    &        0.68(15)      & 0.21(5) \\
    \bottomrule
  \end{tabular}
  \caption{Results for the various topological observables measured at flow time 
$t_0$ on all lattices simulated.\label{tab:results}}  
\end{table}

\begin{wrapfigure}{r}{7.5cm}
  \centering
  \includegraphics{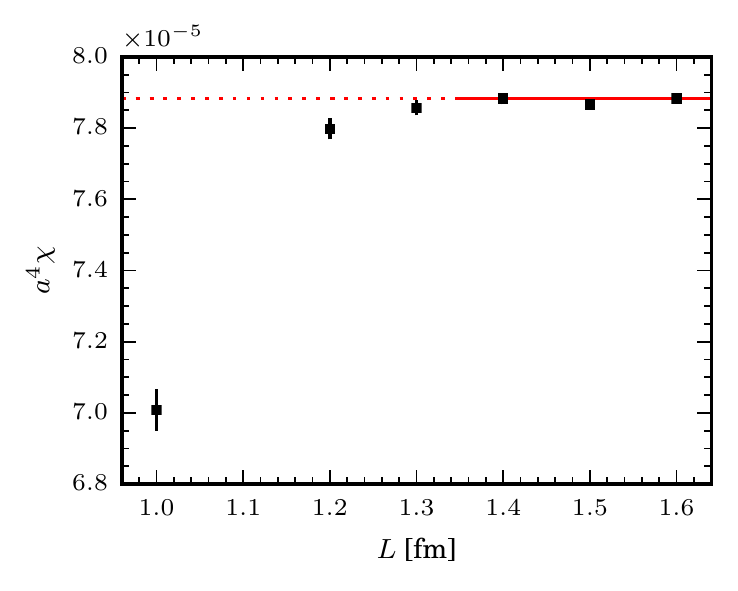}
  \caption{Values of $a^4 \chi$ as a function of $L$ for the series $\{A_1,\dots,F_1\}$.}
  \label{fig:chiV}
\end{wrapfigure}

\begin{figure}[t]
\begin{minipage}{0.35\textwidth}
\includegraphics{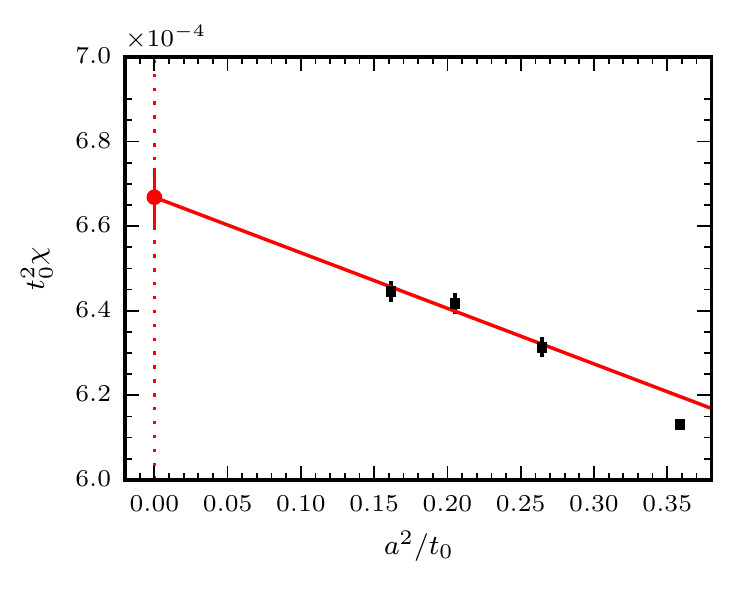}
\end{minipage}
\hspace{20mm}
\begin{minipage}{0.35\textwidth}
\includegraphics{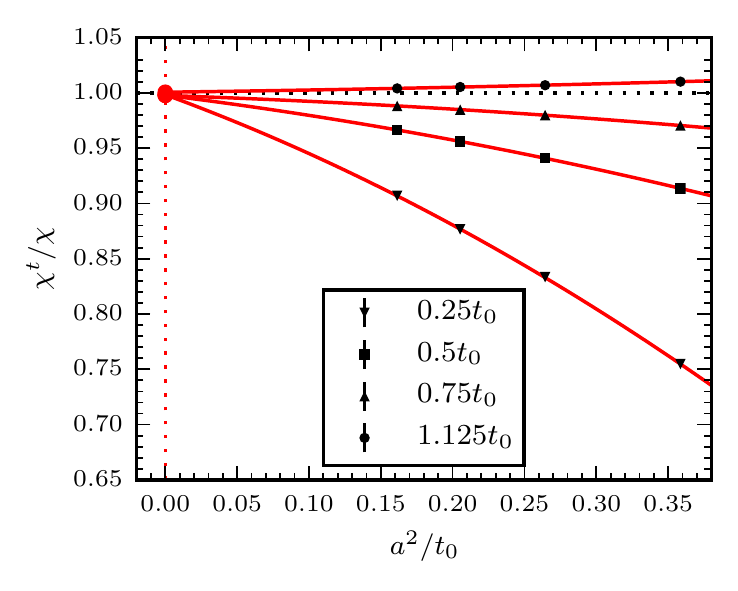}
\end{minipage}
\caption{Right: the dimensionless quantity $t_0^2\,\chi$ as a function of $a^2/t_0$, 
and its extrapolation to the continuum limit. Left: the ratio 
$\chi^t/\chi$ (errors are smaller than symbols) as a function of $a^2/t_0$ for several values of $t$, and its 
extrapolation to the continuum limit.}
\label{fig:chiC}
\end{figure}

\noindent In \figref{fig:chiV} we show the values of the topological susceptibility 
$\chi=\langle Q^2\rangle/V$ from the lattices $\{A_1,\dots,F_1\}$ as a 
function of the linear extension of the lattice. For $L\geq 1.4$~fm, finite-size effects turn out 
to be below our target statistical error of approximatively $0.5\%$. The continuum value
of  $t_0^2 \chi$ can thus be obtained by extrapolating the results 
from the lattices $\{D_1,\dots, D_4\}$, see left plot of \figref{fig:chiC}.
The Symanzik effective theory analysis predicts discretization errors to start at $O(a^2)$, and  
indeed the four data points are compatible with a linear behavior in $a^2$. A linear fit of all 
of them
gives as intercept $t_0^2\,\chi=\num{6.75(4) e-4}$ with a significance of 
$\chi^2/{\rm dof}=1.26$. A quadratic fit gives $t_0^2\,\chi=6.49(18) \cdot 10^{-4}$
with $\chi^2/{\rm dof}=0.38$, and with a coefficient of the quadratic term
compatible with zero within the statistical errors. 
By restricting the linear fit to the three points at the finer lattice spacings, we obtain 
 \begin{equation}
\label{eq:t2chi_continuum_limit}
  t_0^2\, \chi  = \num{6.67(7) e-4}\,, 
\end{equation}
with $\chi^2/{\rm dof}=0.88$, 
which is our best result for this quantity. 
It is five times more precise than the 
determination which uses the Neuberger's definition  of the 
topological charge~\cite{DelDebbio:2004ns}.

The cumulants of the topological charge are expected to be $t$-independent in the 
continuum limit. In the right plot of  Figure~\ref{fig:chiC} we show the topological
susceptibility computed at various flow-times normalized to its value 
at $t_0$. The data points have statistical errors which range from  
$0.1$ to $1$ permille due to the correlation between the numerator and the 
denominator. 
At finite lattice spacing discretization effects are clearly visible, and 
they depend on $t$. When each set of data is extrapolated to the continuum limit
with a quadratic function in $a^2/t_0$, the intercepts are all compatible with 1 within 
the statistical errors which, depending on $t$, range from 0.5 to 5 permille.
We can also compare our result in Eq.~(\ref{eq:t2chi_continuum_limit}) with the one
obtained almost 10 years ago with the Neuberger's definition of the topological 
charge~\cite{DelDebbio:2004ns}. If we use Eqs.~(\ref{eq:t0r0}) and (\ref{eq:t2chi_continuum_limit}), 
we obtain 
\be
r_0^4 \, \chi = 0.0544(18)\, ,  
\ee
which differs by less than 1.5 standard deviations\footnote{This value takes into account the fact 
that the same determination of $r_0$ is used in the two computations.} from the result in Eq.~(11)
of Ref.~\cite{DelDebbio:2004ns}. It is interesting to note that after 
ten years from the first computation of $\chi$ in the continuum limit~\cite{DelDebbio:2004ns}, 
we moved from an unsolved 
problem to a universality test at the permille level\footnote{A first test of universality
for $\chi$ was already presented in Ref.~\cite{Luscher:2010ik} with statistical 
errors more than one order of magnitude larger than those obtained here.
Results with similar large statistical errors were recently obtained 
in Ref.~\cite{Cichy:2015jra}.}. 

By using the result in
Eq.~(\ref{eq:t0fm_continuum_limit}), the value of $\chi$ in physical units 
is given by\footnote{Note that in Ref.~\cite{DelDebbio:2004ns}, $F_K=113.1$~MeV was
used to set the scale in physical units.
}
\be
\chi = (180.5(5)(43)~{\rm MeV})^4\; . 
\ee
where the first error is statistical from Eq.~(\ref{eq:t2chi_continuum_limit}), while 
the second is the one from the uncertainty in the scale in 
Eq.~(\ref{eq:t0fm_continuum_limit}). If we use the physical value of $t_0$ determined
in QCD with $N_f=2$ and $N_f=2+1$ flavours~\cite{Borsanyi:2012zs,Bruno:2013gha},
we obtain a value of $\chi$ in physical units which differs (downwards) by
$10$--$20$\% per linear dimension. This is the size of the ambiguity
which is expected when results of the Yang--Mills theory are expressed
in physical units.

\subsection{The ratio \texorpdfstring{$R$}{R}}
\begin{figure}[t]
\begin{minipage}{0.35\textwidth}
\includegraphics{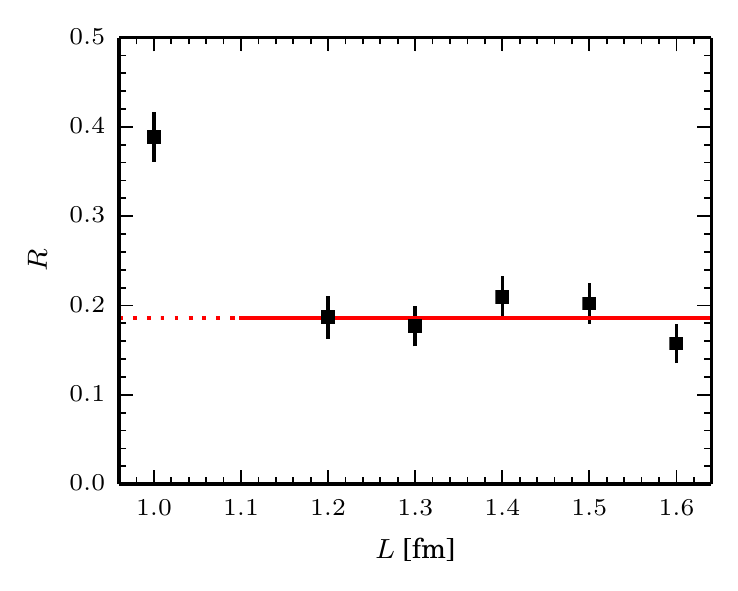}
\end{minipage}
\hspace{20mm}
\begin{minipage}{0.35\textwidth}
\includegraphics{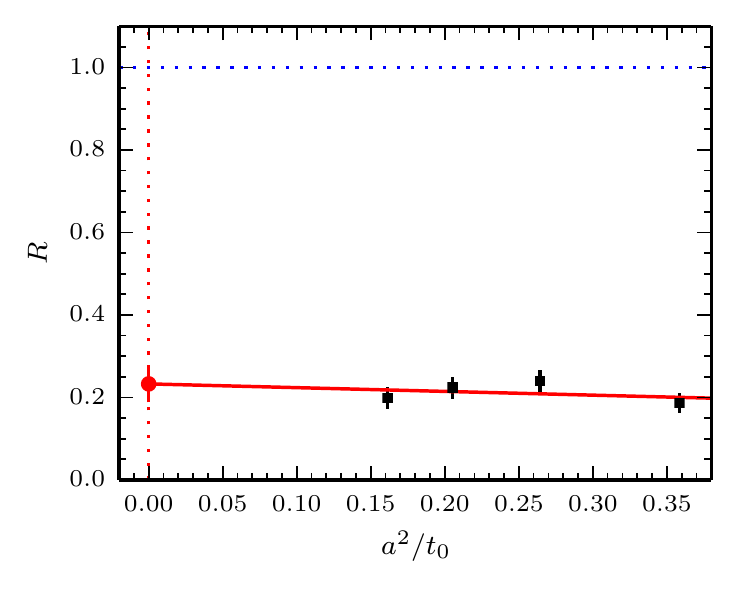}
\end{minipage}
\caption{Left: values of $R$ at flow-time $t_0$ versus $L$ for the series $\{A_1,\dots,F_1\}$. 
Right: the quantity $R$ as a function of $a^2/t_0$ and its extrapolation 
to the continuum limit; the dotted blue line is the dilute instanton gas model prediction $R=1$.}
\label{fig:RC}
\end{figure}

The values of $R=\langle Q^4\rangle_c/\langle Q^2\rangle$
from the lattices $\{A_1,\dots,F_1\}$ are shown in the left plot of 
\figref{fig:RC} as a function of $L$. Since our target statistical error is 
approximatively $10\%$, a linear extension of $L\geq 1.2$~fm is enough for 
finite-size effects to be within errors. Given the increase with $L^8$ of 
the computational cost of $R$, we have chosen to determine its continuum limit  
by extrapolating the data from the lattices $\{B_1,\dots, B_4\}$, see left plot of \figref{fig:RC}.
Also in this case the Symanzik effective theory analysis predicts discretization errors to start 
at $O(a^2)$, and indeed the four data points are compatible with a linear behavior in $a^2$. 
A fit to a constant of all of them gives $R=0.210(13)$ with a significance of 
$\chi^2/{\rm dof}=0.83$. A linear fit in $a^2/t_0$ gives 
\begin{equation}
\label{eq:R_continuum_limit}
  R  = 0.233(45)\, ,   
\end{equation}
which is our best result for this quantity. The significance of the fit 
is $\chi^2/{\rm dof}=1.1$, and the slope is compatible with zero. 

The value in Eq.~(\ref{eq:R_continuum_limit}) is compatible 
with the one obtained with the Neuberger's definition in Ref.~\cite{Giusti:2007tu}, albeit 
with an error 2.5 times smaller. It is also relevant to note that a systematic study of 
finite-size effects was not carried out in Ref.~\cite{Giusti:2007tu}, and finite-size effects were 
estimated and added to the final error.

\section{Conclusions}
The $\theta$-dependence of the vacuum energy, or equivalently the functional 
form of the topological charge distribution, is a distinctive feature of the 
ensemble of gauge configurations that dominate the path integral of a 
Yang-Mills theory. The value of $R=0.233(45)$ in Eq.~(\ref{eq:R_continuum_limit}) 
rules out the $\theta$-behavior predicted by the dilute instanton gas model. Its large 
distance from 1 implies that, in the ensemble of gauge configurations 
that dominate the path integral, the fluctuations of the topological charge 
are of quantum {\it non-perturbative} nature. The large $N_c$ expansion does not 
provide a sharp prediction for $R$. Its small value, however, is compatible with 
being a quantity suppressed as $1/N_c^2$ in the limit of large number of colors 
$N_c$. The value of $R$ found here is related via the Witten–Veneziano mechanism to the
leading anomalous contribution to the $\eta'$--$\eta'$ elastic scattering
amplitude in QCD. It is one of the low-energy constants which enter the
effective theory of QCD when its Green functions are expanded simultaneously
in powers of momenta, quark masses and $1/N_c$.

The Yang--Mills gradient flow is an extremely powerful tool for studying 
the topological properties of the theory. It provides a reference 
scale and a sensible definition of the topological charge which 
are cheap to be computed numerically. With a modest numerical 
effort by today standards, it allowed us to compute the dimensionless 
ratio $t_0^2 \chi=\num{6.67(7) e-4}$ with a relative error 
of roughly $1\%$ in the continuum limit, i.e. five times smaller than 
the one of the previous reference computation with the Neuberger's definition.
The Yang--Mills gradient flow is clearly an interesting tool to study the 
topological properties of the Yang--Mills vacuum as a function of $N_c$.

As proven in this paper, in the continuum limit the cumulants of topological
charge defined by the Yang--Mills gradient flow coincide with those of the
universal definition appearing in the chiral
Ward identities. This in turn implies that this definition of the
topological charge is the correct one
for studying the $\theta$-dependence of the vacuum of QCD at zero and
non-zero temperature. If computed in thermal (full) QCD, its cumulants can
be directly related, for instance, to the axion dynamics without further
renormalization.

\section{Acknowledgments}
We thank M.~L\"uscher for several illuminating discussions which were 
instrumental to derive the results in sections \ref{sec:cont} and
\ref{sec:Qlat}, and for
giving us many suggestions to improve the first version of the paper.
Simulations have been performed on the PC-cluster Turing and Wilson
at Milano-Bicocca, and on Tramontana
at Pisa. We thank these institutions for the computer resources and 
the technical support. G.P.E. and L.G.~acknowledge partial support by 
the MIUR-PRIN contract 20093BMNNPR and by the INFN SUMA project. 

\appendix

\section{Definition and conventions\label{ap:conventions}}
The Lie algebra of the SU(3) group may be identified with the linear space of all
hermitian traceless $3\times 3$ matrices. In the basis 
$T^a$, $a=1 \dots 8$, with 
\be
\tr[\,T^a] = 0 \; ,\quad T^{a\dagger} = T^a\; ,
\ee
the elements of the algebra are linear combinations
of them with real coefficients. The structure constants $f^{abc}$ in the 
commutator relation  
\be
[T^a,T^b]   =  i f^{abc}T^c 
\ee
are real and totally anti-symmetric in the indices if the normalization
condition
\be
\tr[\,T^a T^b] = \frac{1}{2}\delta^{ab}
\ee
is imposed.\\[-0.25cm] 

For the SU(3) Yang--Mills theory the standard Wilson plaquette 
action is given by 
\be
S[U] = \frac{\beta}{2}\, a^4\sum_{x} \sum_{\mu,\nu} 
\left[1 - \frac{1}{3}{\rm Re}\tr\Big\{U_{\mu\nu}(x)\Big\}\right]\; ,
\ee
where the trace is over the color index, $\beta=6/g_0^2$ with $g_0$ 
the bare coupling constant, $a$ is the lattice spacing, and 
the plaquette is defined as a function 
of the gauge links $U_\mu(x)$ as
\be\label{eq:placst}
U_{\mu\nu}(x) = U_\mu(x)\, U_\nu(x+ a\hat \mu)\, U^\dagger_\mu(x + a\hat \nu)\,
             U^\dagger_\nu(x)\; ,  
\ee
with $\mu,\,\nu=0,\dots,3$, $\hat \mu$ is the unit vector along the 
direction $\mu$ and $x$ is the space-time coordinate.\\[-0.25cm] 

The Neuberger-Dirac operator is defined as~\cite{Neuberger:1997bg}
\bea\label{eq:helpN}
D & = & \frac{1}{\bar a}\left\{1+\gamma_{5}\,\mbox{sign}(H)\right\},\nonumber\\
H & = & \gamma_{5}\left(a D_{\rm w} -1-s\right)\;, \qquad
  \bar a =\frac{a}{1+s}\;,
\eea
where $s$ is a real parameter in the range $|s|<1$, and 
and $D_{\rm w}$ is the the Wilson-Dirac operator. It is 
defined as
\be
  D_{\rm w}=\frac{1}{2}\left\{
  \gamma_{\mu}(\nabla^*_{\mu}+\nabla_{\mu})-a\nabla^*_{\mu}\nabla_{\mu}\right\}\; ,
\ee
where
\bea
  \nabla_{\mu}f(x) & = & \frac{1}{a}
  \left\{U_\mu(x)f(x+a\hat{\mu})-f(x)\right\}\, , \\
  \nabla^*_{\mu}f(x)& = & \frac{1}{a}
  \left\{f(x)-U^\dagger_\mu(x-a\hat{\mu}) f(x-a\hat{\mu})\right\}
\eea
are the gauge-covariant forward and backward difference
operators. The Neuberger-Dirac operator satisfies the 
GW relation
\be
\gamma_5 D + D \gamma_5 = \bar a D \gamma_5 D \; . 
\ee
The link differential operators acting on functions $f(U)$ of the 
gauge field are
\begin{equation}
\label{eq:link_differential_op}
  \partial^a_{x,\mu} f(U) = \dv{s} \eval{f(e^{-i s X}U)}_{s=0}, \quad X_\nu(y) = 
\begin{cases} T^a & 
\qif* (y,\nu) = (x,\mu) \\
0   & \text{otherwise}
\end{cases}\, .
\end{equation}
While these depend on the choice of the generators $T^a$, the combination
\begin{equation}
\partial_{x,\mu} f(U) = T^a \partial^a_{x,\mu} f(U)
\end{equation}
can be shown to be basis-independent.

\section{Runge--Kutta--Munthe-Kaas integrators}
\label{ap:runge_kutta}
Consider an ordinary differential equation
\begin{equation}
\label{eq:ODE_G}
  \dot{y} = f(y)y\, , \qquad y(0) = y_0\; ,
\end{equation}
where $y\in G$ for some Lie group $G$ and $f(y)\colon G\to\lalg{g}$, with $\lalg{g}$ being the Lie 
algebra of $G$.
Runge--Kutta--Munthe-Kaas methods ~\cite{Munthe-Kaas:1995,Munthe-Kaas:1998,Munthe-Kaas:1999}
are \emph{structure-preserving} Runge--Kutta methods designed to integrate numerically these equations
on the group manifold, for a general introduction see Ref.~\cite{Hairer:2006}. The starting point is to write the 
solution of~\eqref{eq:ODE_G} as
\begin{equation}
  y(t) = \exp \left\{ v(t) \right\} y(0)\; ,
\end{equation}
and then solve the ordinary differential equation
\begin{equation}
\label{eq:gradient_flow_alg}
  \dot{v} = \dd{\exp_v^{-1}} \left\{ f(y) \right\}, \qquad v(0) = 0\; ,
\end{equation}
where $\dd{\exp_v^{-1}}$ has the series expansion
\begin{equation}
\label{eq:dexpinv_series}
  \dd{\exp_v^{-1}} = \sum_{k=0}^\infty \frac{B_k}{k!} \ad{v}^n = 1 + 
\frac{1}{2} [v,\cdot] + \frac{1}{12} [v,[v,\cdot]] + 
\ldots
\end{equation}
with $B_k$ being the Bernoulli numbers, and $\ad{v}=[v,\cdot]$ the adjoint action. 
Since $v(t)$ takes values in the Lie algebra, the differential 
equation~\eqref{eq:gradient_flow_alg} can be numerically integrated using an ordinary RK method. No extra 
conditions are needed, and any RK method of a given order can be used as a base for a RKMK method of the same 
order. The only complication is given by the operator $\dd{\exp_v^{-1}}$, which can be substituted with its series 
expansion in Eq.~\eqref{eq:dexpinv_series} suitably truncated according to the order of the method. The 
RKMK method of $q^\text{th}$ order with $s$ stages is given by
\begin{equation}
\label{eq:RKMK_s}
  \begin{aligned}
    &\texttt{for } i = 1, 2, \dots, s: \\
    &\qquad\begin{aligned}
      &u_i = \sum_{j=1}^s a_{i,j} \tilde{k}_j \\
      &k_i = h f\{\exp(u_i) y_0\} \\
      &\tilde{k}_i = \text{dexpinv}(u_i, k_i, q)
    \end{aligned} \\
    &v = \sum_{j=1}^s b_j \tilde{k}_j \\
    &y_1 = \exp\{v\} y_0
  \end{aligned}
\end{equation}
where $\text{dexpinv}(u,v,q)$ is the truncated series
\begin{equation}
\label{eq:dexpinv_approx}
  \text{dexpinv}(u,v,q) = \sum_{k=0}^{q-1} \frac{B_k}{k!} \ad{u}^k
\end{equation}
and $a_{i,j}$, $b_i$ are the coefficients of $q^\text{th}$ order $s$-stages RK method.
The fourth order RKMK method that we implemented is obtained starting from the very common 
$4^\text{th}$ order $4$-stages RK method with coefficients, arranged in a Butcher tableau,
\begin{equation}
  \begin{array}{c|cccc}
    0 \\
    \frac{1}{2} & \frac{1}{2} \\
    \frac{1}{2} &       0     & \frac{1}{2} \\
          1     &       0     &       0     &       1 \\
    \hline
                & \frac{1}{6} & \frac{1}{3} & \frac{1}{3} & \frac{1}{6}
  \end{array}
\end{equation}
introduced by Kutta himself. At a first sight this method entails the computation of six different 
commutators of $k_i$ structures. However, it is possible to reduce the number of independent commutators 
needed to only two. As explained in Ref.~\cite{Munthe-Kaas:1999a}, this is due to the fact that, whereas 
the $k_i$ are in general $\order{h}$, some combinations of them are higher order in $h$ and so the corresponding
commutators can be neglected. The resulting integration algorithm is
\begin{equation}
\label{eq:RKMK_4th}
  \begin{aligned}
    u_1 &= 0\, , \quad  k_i = h f\{\exp(u_i) y_0\}\\
     u_2 &= \frac{1}{2}k_1\, ,\\
     u_3 &= \frac{1}{2}k_2 + \frac{1}{8}[k_1, k_2]\, , \\
     u_4 &= k_3\, , \\
     v   &= \frac{1}{6}k_1 + \frac{1}{3}k_2 + \frac{1}{3}k_3 + \frac{1}{6}k_4 - \frac{1}{12}[k_1, k_4]\, , \\
    y_1 &= \exp(v) y_0\; .
  \end{aligned}
\end{equation}
\begin{wrapfigure}{r}{7.5cm}
  \centering
  \includegraphics{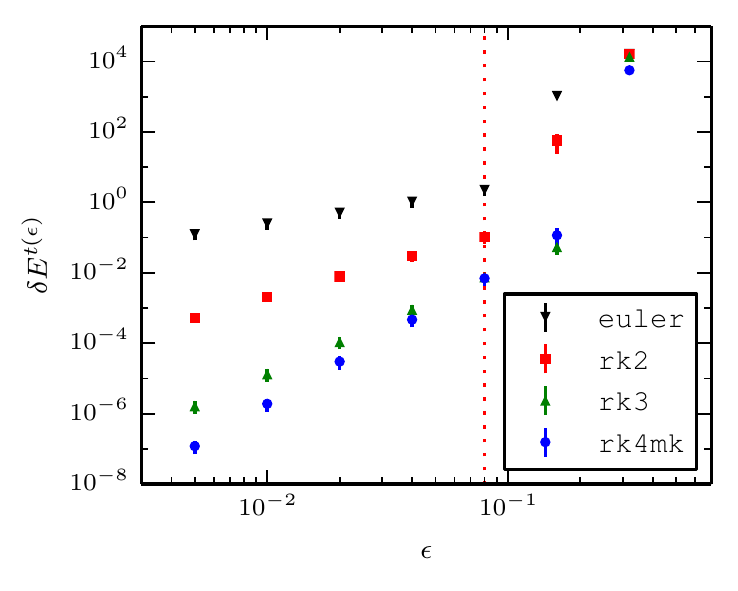}
  \caption{Comparison of the numerical integration methods. The systematic error $\delta E^{t\,(\epsilon)}$, where
           $t=3.2 a^2$, is estimated from $100$ configurations of the lattice $B_1$ by taking the difference 
           between $E^{t\,(\epsilon)}$, evolved with step size $\epsilon$, and 
          $E^{t\,(\epsilon/2)}$, evolved with $\epsilon/2$.}
  \label{fig:RK_comparison}
\vspace{-1.0cm}

\end{wrapfigure}

\noindent Alternative RK methods for integrating \eqref{eq:ODE_G} are given by the Crouch--Grossman 
integrators~\cite{Crouch:1993,Owren:1999}.
They are a special case of so-called \emph{commutator-free} Lie group methods~\cite{Celledoni:2003}. The third order 
algorithm described in Ref.~\cite{Luscher:2010iy} belongs to this class. The conditions which the coefficients 
need to satisfy, order by order, are computable up to arbitrary order~\cite{Owren:2006}. They are given by the 
order conditions for a classical RK method, plus specific extra conditions. At fourth order, however,
we did not find a coefficient scheme with the useful properties of the L\"uscher's 
integrator in terms of exponential reusing.

\subsection{Application to the Yang--Mills gradient flow}
The Yang--Mills gradient flow equation~\eqref{eq:gradient_flow_compact1} can be written as an 
ordinary first-order autonomous differential equation
\begin{equation}
\label{eq:gradient_flow_compact2}
  \dot{V}(t) = Z[V(t)] V(t)\, , \qquad V(0) = V_0\, , 
\end{equation}
where
\begin{equation}
  Z[V(t)] = -g_0^2 \{ \partial_{x,\mu} S[V(t)] \}\; ,
\end{equation}
and the link differential operators are defined in Eq.~\eqref{eq:link_differential_op}. 
The fourth order RKMK method in~\eqref{eq:RKMK_4th} reads
\begin{equation}
\label{eq:RKMK_gradient_flow}
  \begin{aligned}
    W_1              &= V(t)\; , \qquad Z_i = \epsilon Z[W_i]\; , \\
    W_2              &= \exp\left\{ \frac{1}{2}Z_1 \right\} V(t)\; , \\
    W_3              &= \exp\left\{ \frac{1}{2}Z_2 + \frac{1}{8}[Z_1, Z_2] \right\} V(t)\; , \\
    W_4              &= \exp\left\{ Z_3 \right\} V(t)\; , \\
    V(t+a^2\epsilon) &= \exp\left\{ \frac{1}{6}Z_1 + \frac{1}{3}Z_2 + \frac{1}{3}Z_3 + 
\frac{1}{6}Z_4 - \frac{1}{12}[Z_1, Z_4] \right\} V(t)\; .
  \end{aligned}
\end{equation}
This method computes four times the force field $Z[W_i]$ and four times the Lie group exponential. The commutators are economically implemented exploiting structure constants of $\lalg{g}$. Each iteration needs space in memory for one auxiliary gauge field and three $Z_i$ fields. Gauge fields are stored in memory with a full $3\times 3$ complex matrix, which has $18$ real components, for each link. A $Z_i$ field is an element of $\lalg{su}(3)$, which is a $8$ dimensional linear space, for each link. Thus, the method~\eqref{eq:RKMK_gradient_flow} requires space for $(18+3\times 8)\times 4V$ floating point numbers. Each exponential of a $\lalg{g}$-valued combination of $Z[W_i]$ reduce to $4V$ $\lalg{su}(3)$ matrices exponentials, which can be computed economically exploiting the Cayley--Hamilton theorem as described in~\cite{Luscher:2009}.

In the left plot of Figure~\ref{fig:RK_comparison} the RKMK method is compared to lower-order Runge--Kutta methods, 
such as the third order method \texttt{rk3} found in~\cite{Luscher:2010iy}. The comparison is done averaging over
100 configurations at $\beta=5.96$ on a $12^4$ lattice evolved at $t=3.2 a^2$. The \texttt{rk4mk} algorithm scales 
correctly as a fourth-order method. However, the pre-factor appears to be larger, thus the new method is more precise 
with respect to \texttt{rk3} in Ref.~\cite{Luscher:2010iy} for $\epsilon\lesssim 0.1$.

\phantomsection
\bibliographystyle{JHEP}
\bibliography{Literature.bib}

\end{document}